\documentclass[sigconf]{acmart}

\usepackage{amsmath,amsfonts,mathtools}
\usepackage{graphicx}
\usepackage{booktabs}
\usepackage{array}
\usepackage{multirow}
\usepackage{enumitem}
\usepackage{float}
\usepackage{subcaption}
\usepackage{xurl}
\usepackage{acro}
\usepackage{cleveref}
\usepackage{algorithm}
\usepackage{algorithmicx}
\usepackage{algpseudocode}
\usepackage{xcolor}

\hypersetup{
  colorlinks=true,
  linkcolor=blue,
  citecolor=blue,
  urlcolor=blue
}

\newcommand{\vect}[1]{\mathbf{#1}}

\DeclareAcronym{gr}{
  short = GR,
  long  = Generative Recommendation
}
\DeclareAcronym{sid}{
  short = SID,
  long  = Semantic ID
}
\DeclareAcronym{et}{
  short = ET,
  long  = Encoding Table
}
\DeclareAcronym{dt}{
  short = DT,
  long  = Decoding Table
}
\DeclareAcronym{actionbos}{
  short = Action-BOS,
  long  = Action-conditioned Beginning-of-Sequence
}
\DeclareAcronym{bos}{
  short = BOS,
  long  = Beginning-of-Sequence
}
\DeclareAcronym{mmoe}{
  short = MMoE,
  long  = Multi-gate Mixture-of-Experts
}
\DeclareAcronym{sdpa}{
  short = SDPA,
  long  = Scaled Dot-Product Attention
}
\DeclareAcronym{rmsnorm}{
  short = RMSNorm,
  long  = Root Mean Square Layer Normalization
}
\DeclareAcronym{swiglu}{
  short = SwiGLU,
  long  = Swish-Gated Linear Unit
}
\DeclareAcronym{swigluffn}{
  short = SwiGLU-FFN,
  long  = SwiGLU Feed-Forward Network
}
\DeclareAcronym{gatedattn}{
  short = Gated-Attn,
  long  = Gated Attention
}
\DeclareAcronym{ntp}{
  short = NTP,
  long  = Next Token Prediction
}
\DeclareAcronym{hr}{
  short = HR,
  long  = Hit Rate
}
\DeclareAcronym{gt}{
  short = GT,
  long  = Ground Truth
}
\DeclareAcronym{dis}{
  short = DIS,
  long  = Discriminative Recommendation
}
\DeclareAcronym{rq}{
  short = RQ,
  long  = Residual Quantization
}
\DeclareAcronym{gauc}{
  short = GAUC,
  long  = Group Area Under the Curve
}
\DeclareAcronym{rmse}{
  short = RMSE,
  long  = Root Mean Square Error
}
\DeclareAcronym{coa}{
  short = CoA,
  long  = Chain-of-Attribute
}
\DeclareAcronym{cot}{
  short = CoT,
  long  = Chain-of-Thought
}
\DeclareAcronym{cc}{
  short = CDC,
  long  = Conditional Decoding Context
}
\DeclareAcronym{cs}{
  short = CS,
  long  = Content Summary
}
\DeclareAcronym{ge}{
  short = GE,
  long  = Grounding Evidence
}
\DeclareAcronym{bi}{
  short = BI,
  long  = Behavior Instruction
}
\DeclareAcronym{beams}{
  short = BS,
  long  = Beam Search
}
\DeclareAcronym{beamhr}{
  short = BHR,
  long  = Beam Hit Rate
}
\DeclareAcronym{tf}{
  short = TF,
  long  = Teacher Forcing
}
\DeclareAcronym{ca}{
  short = CA,
  long  = Cross-Attention
}
\DeclareAcronym{rdnn}{
  short = R-DNN,
  long  = Rank-DNN
}
\DeclareAcronym{masknet}{
  short = MaskNet,
  long  = MaskNet
}
\DeclareAcronym{senet}{
  short = SENet,
  long  = SENet
}
\DeclareAcronym{dnn}{
  short = DNN,
  long  = Deep Neural Network
}
\DeclareAcronym{adamw}{
  short = AdamW,
  long  = AdamW
}
\DeclareAcronym{validk}{
  short = Valid@K,
  long  = Valid@K
}
\DeclareAcronym{sp}{
  short = TC-BOS,
  long  = Task-Conditioned BOS
}
\DeclareAcronym{llm}{
  short = LLM,
  long  = Large Language Model
}

\copyrightyear{2026}
\acmYear{2026}
\setcopyright{acmcopyright}
\acmConference[Conference '26]{ACM Conference}{Month 2026}{Location}
\acmBooktitle{Proceedings of Conference '26}
\acmPrice{15.00}
\acmDOI{10.1145/XXXXXX.XXXXXX}
\acmISBN{978-1-4503-XXXX-X/26/06}

\begin{document}

\title{UniRec: Bridging the Expressive Gap between Generative and Discriminative Recommendation via Chain-of-Attribute}

\author{
  Ziliang Wang$^\dagger$, Gaoyun Lin$^\dagger$, Xuesi Wang, Shaoqiang Liang, Yili Huang \\
  Weijie Bian$^*$, Li Zhang, Mingchen Cai, Jian Dong, Guanxing Zhang
}

\affiliation{
  \institution{Shopee}
  \country{}
}

\thanks{$^\dagger$Equal contribution. $^*$Corresponding author (jason.bian@shopee.com).}
\begin{abstract}

Generative Recommendation (GR) reframes retrieval and ranking as autoregressive decoding over Semantic IDs (SIDs), unifying the multi-stage pipeline into a single model. Yet a fundamental \emph{expressive gap} persists: discriminative models score items with direct feature access, enabling explicit user--item crossing, whereas GR decodes over compact SID tokens without item-side signal. We formalize this via Bayes' theorem, ranking by $p(y \mid \mathbf{f}, u)$ is equivalent to ranking by $p(\mathbf{f} \mid y, u)$, which factorizes autoregressively over item features, showing that a generative model with full feature access matches its discriminative counterpart, with any practical gap stemming solely from incomplete feature coverage.

We propose \textsc{UniRec} with Chain-of-Attribute (CoA) as its core mechanism. CoA prefixes each SID sequence with structured attribute tokens:category, seller, brand,before decoding the SID, recovering the item-side feature crossing that discriminative models exploit. Since items sharing identical attributes cluster in adjacent SID regions, attribute conditioning yields a measurable per-step entropy reduction $H(s_k \mid s_{<k}, \mathbf{a}) < H(s_k \mid s_{<k})$, narrowing the search space and stabilizing beam search. We further address two deployment challenges: \emph{Capacity-constrained SID} introduces exposure-weighted capacity penalties into residual quantization to suppress token collapse and the Matthew effect; \emph{Conditional Decoding Context (CDC)} combines Task-Conditioned BOS with hash-based Content Summaries to inject scenario signals at each decoding step. A joint RFT and DPO framework aligns the model with business objectives beyond distribution matching.

Experiments show \textsc{UniRec} outperforms the strongest baseline by +22.6\% HR@50 overall and +15.5\% on high-value orders. Deployed on Shopee's e-commerce platform, online A/B tests confirm significant gains in PVCTR (+5.37\%), orders (+4.76\%), and GMV (+5.60\%).

\end{abstract}

\ccsdesc[500]{Information systems~Recommender systems}

\keywords{Generative Recommendation, Semantic ID, E-commerce Recommendation}

\renewcommand{\shortauthors}{Ziliang Wang, 
  Gaoyun Lin,
  Xuesi Wang,
  Shaoqiang Liang,
  Yili Huang,
  Weijie Bian
et al.}

\maketitle

\section{Introduction}

Current recommendation systems employ a multi-stage discriminative 
pipeline---Retrieval $\to$ Pre-ranking $\to$ Ranking $\to$ 
Reranking---that suffers from three well-known limitations: 
inconsistent objectives across stages, sample selection bias where 
ranking models trained on the exposed space fail to generalize to 
the full candidate set, and compounding error propagation through 
the funnel without possibility of downstream correction. Inspired 
by the success of large language models~\cite{gpt, llama, deepseek}, 
Generative Recommendation (GR) reframes 
this as autoregressive decoding over Semantic IDs (SIDs), unifying 
retrieval and ranking into a single model and eliminating 
inter-stage misalignment~\cite{tiger, onerec}.

Despite this architectural advantage, a \textbf{fundamental expressive gap} separates the two 
paradigms. Discriminative models estimate $p(y \mid \vect{f}, u)$ with direct access to 
item feature vectors $\vect{f}$, enabling explicit user--item feature crossing. 
Generative models, by contrast, decode over compact SID tokens without 
conditioning on any item-side signal---the model must 
commit to a generation path before it has seen any item feature. 
This is not merely an engineering limitation; it reflects a \emph{structural asymmetry} in the information available at decoding time. Furthermore, recommendation inherently involves 
\emph{one-to-many} mappings: unlike language modeling where a given 
context largely determines the next token, a single user behavior 
sequence may correspond to multiple valid target items, amplifying 
generation uncertainty and training conflicts that discriminative 
models resolve through explicit feature interaction. This raises 
a fundamental question: \emph{can generative recommendation match the 
expressive power of its discriminative counterpart, and if so, under what conditions?}

We answer this question affirmatively through a Bayesian analysis. 
By Bayes' theorem, ranking items by the discriminative score 
$p(y \mid \vect{f}, u)$ is equivalent to ranking by the generative 
posterior $p(\vect{f} \mid y, u)$, which factorizes autoregressively 
over item features. This equivalence establishes a \emph{theoretical upper bound}: 
a generative model with full feature access is as expressive as its 
discriminative counterpart; any practical gap arises from 
\emph{feature coverage} rather than a fundamental modeling asymmetry. 
The key insight is that SID codes, by compressing rich item semantics 
into compact discrete tokens, inevitably discard structured item 
properties---such as taxonomy, seller, and brand identifiers---that 
discriminative models exploit through explicit feature crossing. 
Recovering these lost signals within the generative trajectory is 
both necessary and sufficient to close the expressive gap.

This insight motivates \textbf{Chain-of-Attribute (CoA)}, the core 
mechanism of our proposed framework \textsc{UniRec}. CoA prefixes each 
SID sequence with coarse-grained attribute tokens, yielding a 
speculate-then-refine generation paradigm: the model first predicts 
structured item attributes (category, seller, brand), then 
conditionally generates the SID given these attributes. 
Since items sharing identical attributes occupy adjacent regions in 
the SID semantic space, attribute conditioning yields a measurable 
per-step conditional entropy reduction---formally, 
$H(s_k \mid s_{<k}, \vect{a}) < H(s_k \mid s_{<k})$---whose 
cumulative effect narrows the effective search space, stabilizes 
beam search trajectories, and attenuates end-to-end decoding errors. 
This mechanism bridges the expressive gap by recovering, within the 
generative decoding trajectory, the item-side feature crossing that 
was previously exclusive to discriminative models.

Beyond CoA, deployment at scale introduces two further 
challenges. {First}, item popularity follows a long-tail 
distribution: even with equal item counts per cluster, a small 
fraction of high-traffic items dominates the exposure load of 
their assigned codebook entries. In our production data, the top 
10\% of $(s_0, s_1)$ combinations capture 87.9\% of total 
exposure---a 2.6$\times$ amplification of mild item-count 
imbalance into extreme traffic concentration. 
{Capacity-constrained SID} addresses this by introducing 
exposure-weighted capacity penalties into residual quantization, 
directly controlling traffic load per codebook entry and 
suppressing the hereditary Matthew effect across SID layers. 
{Second}, a unified model serving multiple recommendation 
scenarios without explicit task conditioning suffers from 
conflicting behavioral distributions that destabilize generation 
and cause errors to compound across the autoregressive chain. 
{Conditional Decoding Context (CDC)} addresses this by 
combining a scenario-conditioned Task-Conditioned BOS with 
hash-based Content Summaries, injecting task-specific and 
combinatorial signals at each decoding step. To further align \textsc{UniRec} 
with business objectives beyond distribution matching, we adopt 
Direct Preference Optimization (DPO)~\cite{dpo} combined with 
Reward-Driven Fine-tuning (RFT) that reweights training samples 
by continuous business value estimates.

Our main contributions are as follows:
\begin{itemize}
    \item \textbf{Theoretical Characterization of the Expressive Gap}: 
    We provide the first information-theoretic analysis showing that 
    the expressive gap between generative and discriminative 
    recommendation arises from feature coverage, not modeling 
    asymmetry. Via Bayes' theorem, we establish a theoretical upper 
    bound and prove that a generative model with full feature access 
    matches the discriminative ranking.

    \item \textbf{Chain-of-Attribute (CoA)}: 
    Motivated by the theoretical analysis, we propose CoA as a 
    principled mechanism to bridge the expressive gap. By 
    pre-generating structured item attributes before SID decoding, 
    CoA recovers item-side feature crossing within the generative 
    trajectory, yielding measurable per-step entropy reduction 
    $H(s_k \mid s_{<k}, \vect{a}) < H(s_k \mid s_{<k})$ and 
    end-to-end error attenuation.

    \item \textbf{Capacity-constrained SID and CDC}: 
    We address SID distribution collapse through exposure-weighted 
    residual quantization, and multi-scenario error accumulation 
    through Task-Conditioned BOS and combinatorial Content 
    Summaries---two system-level contributions essential for 
    large-scale deployment.

    \item \textbf{Comprehensive Empirical Validation}: 
    Extensive offline experiments show that \textsc{UniRec} 
    outperforms the strongest baseline by +22.6\% relative HR@50 
    on all samples and +15.5\% on high-value order samples. Deployed on Shopee's e-commerce platform, online A/B tests confirm significant gains in PVCTR (+5.37\%), orders (+4.76\%), and GMV (+5.60\%).
\end{itemize}

\section{Related Work}

\paragraph{Industrial-Scale Discriminative Recommendation}
Traditional recommendation systems adopt a multi-stage 
discriminative pipeline---retrieval, pre-ranking, ranking, and 
reranking---where each stage optimizes a local objective. While 
effective at scale, this architecture suffers from objective 
misalignment across stages, sample selection bias, and error 
propagation through the funnel. To raise the performance ceiling 
within this paradigm, HSTU~\cite{hstu} proposes sparse sequence 
transduction for high-throughput sequential modeling, and 
RankMixer~\cite{rankmixer} improves model flops utilization 
through architectural redesign. MTGR~\cite{mtgr} combines 
generative sequence modeling with traditional feature-interaction 
modules for industrial-grade optimization. However, because the 
underlying multi-stage structure is preserved, issues of 
misalignment and bias remain fundamentally unresolved.

\paragraph{Generative Recommendation and Semantic IDs}
GR reframes candidate generation as 
autoregressive decoding over discrete item representations, 
unifying retrieval and ranking into a single model. TIGER~\cite{tiger} 
pioneered this direction by mapping items to hierarchical semantic 
tokens via RQ-VAE and predicting token sequences autoregressively. 
Subsequent work has extended this foundation in multiple 
directions: OneRec~\cite{onerec} unifies the multi-stage pipeline 
into an end-to-end generative framework with iterative preference 
alignment; OneMall~\cite{onemall} applies a similar paradigm to 
e-commerce; OneSearch~\cite{onesearch} and 
OneSearch-V2~\cite{onesearchv2} extend generative 
retrieval to e-commerce search with latent reasoning and 
DPO-based alignment. On the tokenization side, 
COINS~\cite{coins} and FORGE~\cite{forge} study the quality and 
collision properties of semantic ID construction, establishing 
that token usage collapse and invalid combinations directly affect 
decoding availability.

GRACE~\cite{grace} introduces Chain-of-Thought tokenization that 
prepends explicit product knowledge graph attributes before 
semantic tokens during decoding, empirically demonstrating that 
attribute conditioning improves generation quality. While GRACE 
and our CoA share the high-level intuition of prefixing attributes 
before SID tokens, they differ in both motivation and scope. 
GRACE treats attribute prepending as an empirical design choice 
for enriching token-level reasoning without theoretical 
justification. CoA, by contrast, is grounded in a formal 
information-theoretic analysis that establishes a theoretical 
upper bound between generative and discriminative recommendation 
via Bayes' theorem, and proves that attribute conditioning yields 
a measurable per-step entropy reduction---providing the first 
theoretical characterization of why attribute prefixing narrows 
the expressive gap in GR.

\paragraph{Preference Alignment for Recommendation}
Aligning GR models with business objectives 
beyond NTP has emerged as a critical challenge 
for deployment at scale. DPO~\cite{dpo} simplifies preference alignment by directly 
optimizing a contrastive objective without a separate reward 
model, and has been applied to recommendation to bridge the gap 
between generative training and user preference 
signals~\cite{onerec, onesearch}. OneSearch-V2~\cite{onesearchv2} 
further demonstrates that DPO-based alignment outperforms reward 
model-based approaches in both flexibility and alignment quality 
for generative retrieval, providing empirical evidence that direct 
preference optimization is more effective than two-stage 
reward-then-policy pipelines. Reward-weighted regression 
approaches~\cite{rwr} reweight training samples by estimated 
business value, providing a lightweight alternative to full 
reinforcement learning. \textsc{UniRec} combines both paradigms 
in a unified framework: following the DPO-first principle 
validated by OneSearch-V2, we adopt DPO to inject discrete 
behavioral preference signals constructed directly from online 
traffic without human annotation, while RFT 
reformulates the NTP objective with continuous business value 
reweighting to capture complementary reward signals.

\section{Method}

\subsection{Overview}

\begin{figure*}[h]
  \centering
  \includegraphics[width=1.0\linewidth]{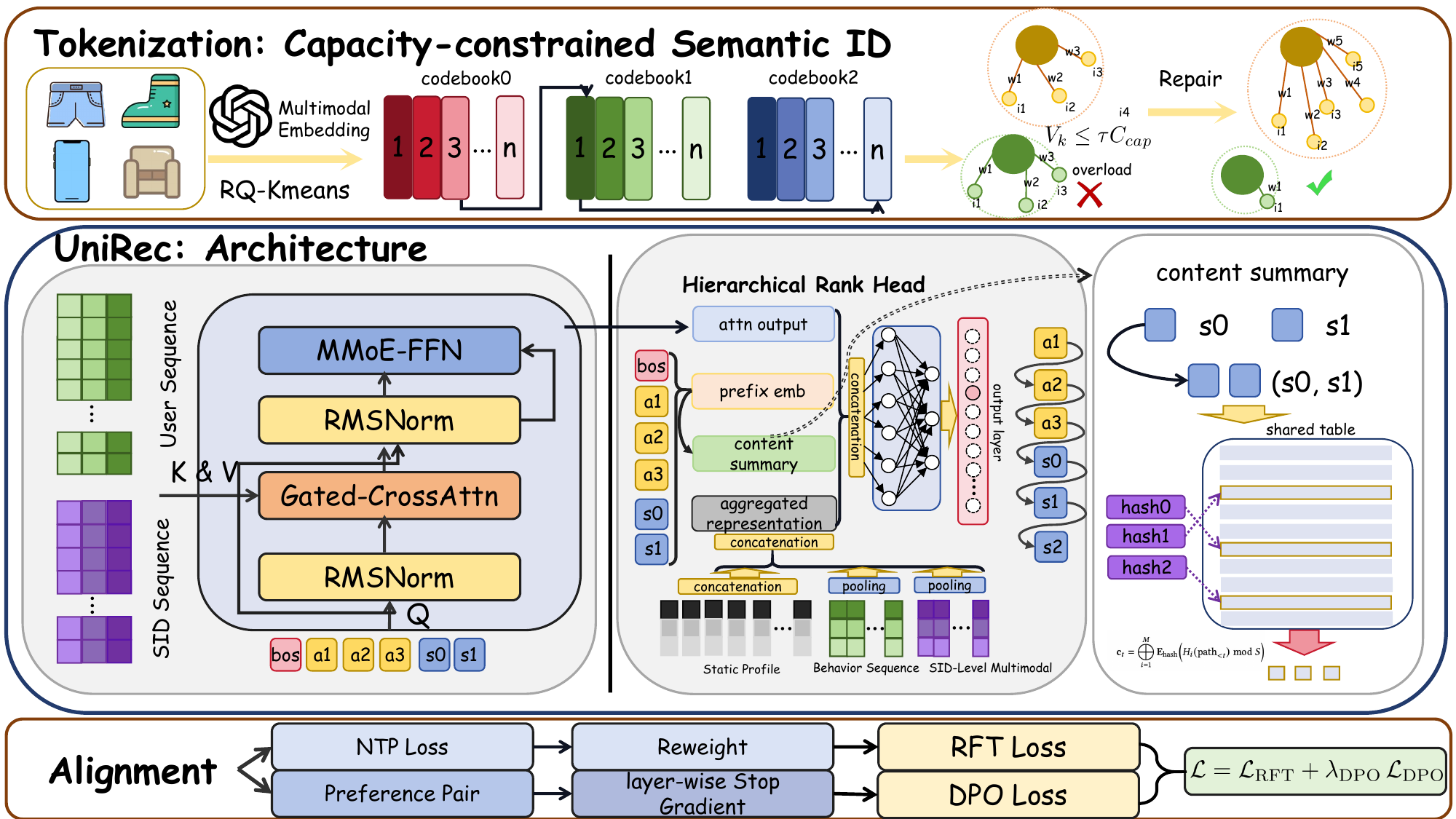}
  \caption{Overview of \textsc{UniRec} architecture.}
  \label{fig:cag_overview}
\end{figure*}

\textsc{UniRec} integrates candidate generation and ranking into a unified autoregressive decoding framework over hierarchical token sequences, where each item is represented by multi-layer semantic tokens. To mitigate the hereditary Matthew effect in hierarchical representations, we propose {Capacity-constrained SID} (\cref{sec:sid}) that enforces balanced token distribution across layers. To enhance generation quality, we introduce {CoA} (\cref{sec:coa}) that speculatively generates item attributes before SID tokens, and {CDC} (\cref{sec:cdc}) that injects task conditioning and combinatorial feature interactions into the decoding process. Our model adopts a Decoder-Only architecture with Cross-Attention to user behavior sequences, where per-step Rank Heads generate token distributions (\cref{sec:model}), followed by RFT and DPO (\cref{sec:alignment}) that align the model with business objectives through value-based sample reweighting. The overall pipeline is illustrated in Figure~\ref{fig:cag_overview}.

\subsection{Capacity-constrained Semantic ID}
\label{sec:sid}

Based on the hierarchical SID framework, we map items to discrete token 
sequences via residual quantization~\cite{tiger}. Prior work on 
balanced quantization~\cite{onerec} enforces uniform \emph{item-count} 
constraints during clustering. However, item popularity in recommendation 
follows a long-tail distribution: even with equal item counts per cluster, 
a small fraction of high-traffic items can dominate the exposure load of 
their assigned codebook entries, leaving most tokens severely underutilized 
in practice. This exposure imbalance gives rise to a \emph{hereditary Matthew effect}: 
high-traffic token combinations accumulate disproportionate exposure in 
training data, causing the model to repeatedly generate the same narrow set 
of token paths during beam search while the vast majority of codebook entries 
remain marginalized.

\begin{figure}[h]
  \centering
  \includegraphics[width=1.0\linewidth]{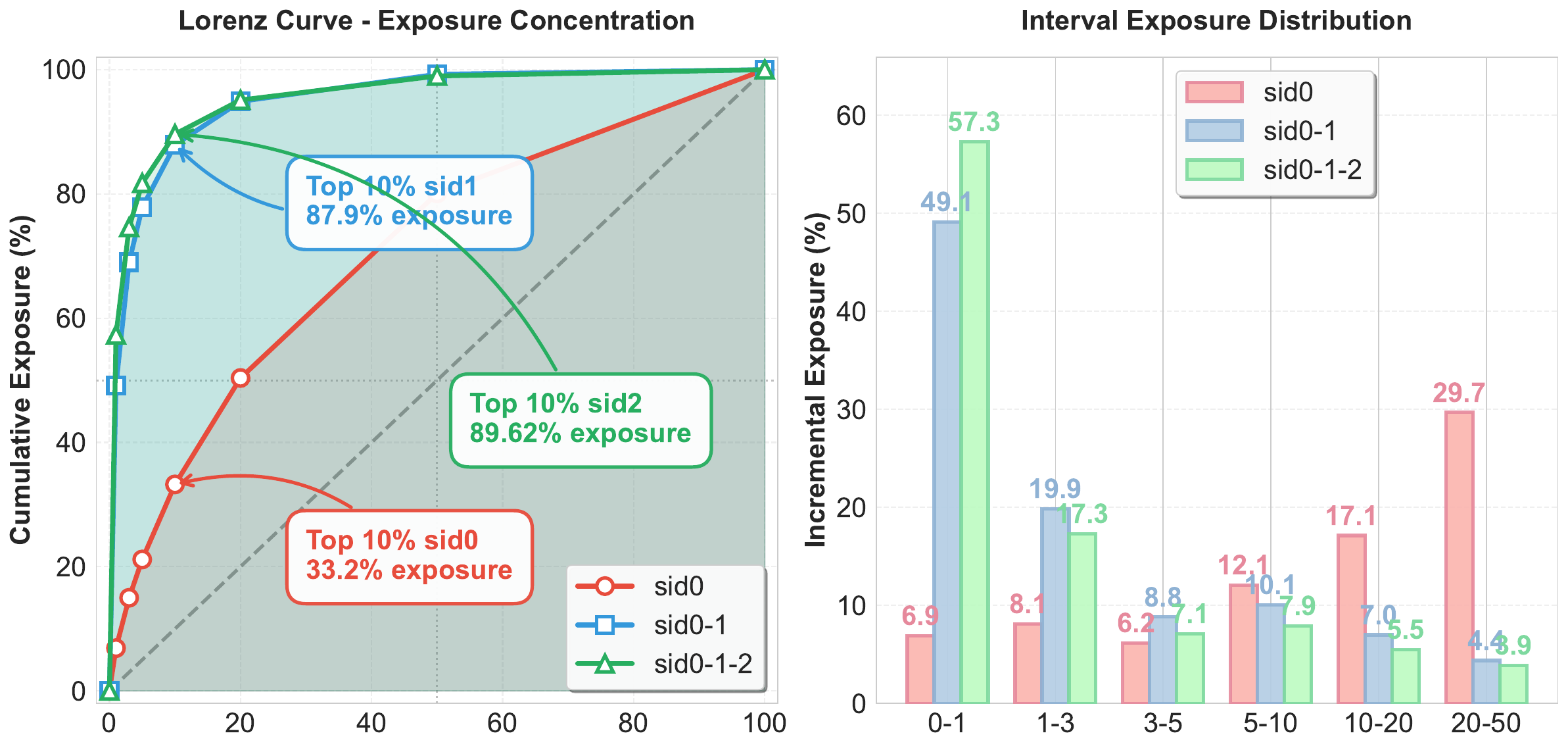}
  \caption{Exposure concentration across SID layers under standard 
  RQ-$K$Means~\cite{onerec}.}
  \label{fig:matthew_dist}
\end{figure}

To quantify this phenomenon, we analyze the exposure distribution using 
production traffic data. For the first layer $s_0$, we count how many times 
each token appears in user impressions, rank tokens by exposure volume, and 
compute the fraction of total impressions captured by the top-$k$\% tokens. 
For deeper layers, we measure exposure at the joint-combination level: for 
$s_1$, we treat each $(s_0, s_1)$ pair as a unique combination and compute 
the cumulative exposure share of the top-$k$\% combinations globally; 
similarly for $s_2$, using $(s_0, s_1, s_2)$ triplets. As shown in 
Figure~\ref{fig:matthew_dist}, the top 10\% of $s_0$ tokens account for 
33.24\% of total exposure---a moderate skew. In stark contrast, the top 10\% 
of $(s_0, s_1)$ combinations capture 87.90\% of exposure, and the top 10\% 
of $(s_0, s_1, s_2)$ combinations capture 89.62\%. This $2.6\times$ 
amplification from $s_0$ to $s_1$ confirms that mild item-count imbalance 
is transformed into extreme exposure concentration at the combination level, 
where only a narrow set of high-frequency token paths dominates nearly all 
traffic.

Capacity-constrained SID addresses hereditary concentration by enforcing 
balanced load distribution in residual quantization clustering. Let the 
clustering samples at layer $l$ be $\{x_i\}_{i=1}^{N}$, where $N$ is the 
number of items, each with exposure weight $w_i>0$ (e.g., historical 
impression count). We denote by $z_i \in \{1,\dots,K\}$ the cluster 
assignment of sample $i$, and by $\boldsymbol{\mu}_k \in 
\mathbb{R}^{d_{\text{emb}}}$ the center of the $k$-th cluster. Define the 
exposure load (volume) of the $k$-th cluster as:
\begin{equation}
  V_k = \sum_{i: z_i = k} w_i.
  \label{eq:cap_load}
\end{equation}
Given capacity cap $C_{cap}>0$ and tolerance $\tau\ge 1$, the constrained 
clustering problem is:
\begin{equation}
\begin{aligned}
  \min_{\{z_i\}, \{\boldsymbol{\mu}_k\}} \quad
  & \sum_{i=1}^{N} \|x_i - \boldsymbol{\mu}_{z_i}\|_2^2, \\
  \text{s.t.} \quad
  & V_k \le \tau C_{cap}, \quad \forall k \in \{1, \dots, K\},
\end{aligned}
  \label{eq:cap_kmeans_hard}
\end{equation}
where $\boldsymbol{\mu}_{z_i}$ denotes the center of the cluster to which 
sample $i$ is assigned. The capacity reference is set to $C_{cap} = 
\frac{1}{K}\sum_{i=1}^N w_i$, i.e., the mean exposure load per cluster. 
This hard-constrained problem is NP-hard in general. We adopt a two-phase 
greedy strategy: first assign each sample to its nearest cluster center to 
minimize reconstruction error, then repair overloaded clusters by reassigning 
excess samples to the nearest under-capacity cluster. 
Algorithm~\ref{alg:cc_rq} summarizes the full procedure.

\begin{algorithm}[ht!]
\caption{Capacity-Constrained Residual Quantization}
\label{alg:cc_rq}
\begin{algorithmic}[1]
\Require Item embeddings $\{\vect{e}(x_i)\}_{i=1}^N$, exposure weights 
  $\{w_i\}$, layers $L$, clusters $K$, tolerance $\tau$
\Ensure Semantic IDs $\{s_0(x_i),\dots,s_{L-1}(x_i)\}_{i=1}^N$
\For{$l = 0$ \textbf{to} $L-1$}
  \State $\vect{r}_i \gets \vect{e}(x_i) - \sum_{j<l}
    \boldsymbol{\mu}_{j,s_j(x_i)}$ 
    \Comment{Compute residual; $\vect{r}_i = \vect{e}(x_i)$ if $l=0$}
  \State Initialize $\{\boldsymbol{\mu}_k\}_{k=1}^K$ via $K$-Means++; 
    set $C_{cap} \gets \frac{1}{K}\sum_{i=1}^N w_i$
  \Repeat
    \State $z_i \gets \arg\min_k \|\vect{r}_i - \boldsymbol{\mu}_k\|^2$ 
      for all $i$; compute $V_k \gets \sum_{i: z_i=k} w_i$
    \For{$k$ with $V_k > \tau C_{cap}$} 
      \Comment{Repair overloaded clusters}
      \For{$i$ with $z_i = k$}
        \State $z_i \gets \arg\min_{k': V_{k'} + w_i \le \tau C_{cap}} 
          \|\vect{r}_i - \boldsymbol{\mu}_{k'}\|^2$
        \State Update $V_k \gets V_k - w_i$; $V_{k'} \gets V_{k'} + w_i$
      \EndFor
    \EndFor
    \State $\boldsymbol{\mu}_k \gets \text{mean}\{\vect{r}_i : z_i=k\}$; 
      $\mathcal{J} \gets \frac{1}{N}\sum_i \|\vect{r}_i - 
      \boldsymbol{\mu}_{z_i}\|^2$
  \Until{$|\Delta\mathcal{J}| < \epsilon$}
  \State $s_l(x_i) \gets z_i$, $\boldsymbol{\mu}_{l,k} \gets \boldsymbol{\mu}_k$
\EndFor
\end{algorithmic}
\end{algorithm}

\subsection{Chain-of-Attribute}
\label{sec:coa}

\paragraph{From Discriminative to Generative: A Theoretical Upper Bound.}
A common concern about GR is that, without direct access to
target-side item features during generation, the model's expressive ceiling is
fundamentally lower than that of discriminative ranking models, which score
items by $p(y \mid \vect{f}, u)$ with direct access to item features
$\vect{f}$, whereas generative models factorize over a compact SID
sequence $\prod_{l=0}^{L-1} p(s_l \mid s_{<l}, u)$ without explicit feature
conditioning. We argue this gap is not a fundamental modeling asymmetry, but
rather a consequence of feature coverage. Let $\vect{f} = (f_1, \dots, f_n)$
denote the feature representation of an item, $u$ the user context, and $y$
the engagement label. By Bayes' theorem,
\begin{equation}
  p(y \mid \vect{f}, u) \;\propto\; p(\vect{f} \mid y, u)\,p(y \mid u),
  \label{eq:bayes_disc_gen}
\end{equation}
and since $p(y \mid u)$ is independent of $\vect{f}$, ranking by the
discriminative score is \emph{equivalent} to ranking by the generative
posterior $p(\vect{f} \mid y, u)$. Expanding via the chain rule,
\begin{equation}
  p(\vect{f} \mid y, u) = \prod_{k=1}^{n} p(f_k \mid f_{<k}, y, u),
  \label{eq:chain_rule_features}
\end{equation}
which corresponds precisely to autoregressive decoding over item features.
This establishes a \emph{theoretical upper bound}: a generative model with
full access to item features $\vect{f}$ is expressively equivalent to its
discriminative counterpart. Any practical gap therefore arises from
\emph{approximation capacity} or \emph{feature coverage}, not from a
fundamental modeling asymmetry.

\paragraph{Chain-of-Attribute as Explicit Feature Compensation.}
Fully sequential decoding over all item features is latency-prohibitive at
recommendation scale. GR systems therefore employ hierarchical SID
tokens to compress rich item semantics into a compact code, replacing the
full factorization of \cref{eq:chain_rule_features} with
$\prod_{l=0}^{L-1} p(s_l \mid s_{<l}, u)$. This compression is inherently
\emph{lossy}: structured item properties that discriminative models 
exploit through explicit feature crossing---e.g., taxonomy, seller, 
and brand identifiers---are folded into the SID code and become 
latent during decoding, causing the generative model to fall short 
of the discriminative feature coverage upper bound established above.

\textsc{UniRec} addresses this through the CoA mechanism, which
prefixes the SID sequence with $m$ coarse-grained attribute tokens
$\vect{a} = [\text{attr}_1, \dots, \text{attr}_m]$, yielding the
factorization:
\begin{equation}
  p(\mathbf{s} \mid u) = p(\vect{a} \mid u) \cdot \prod_{l=0}^{L-1}
  p(s_l \mid \vect{a}, s_{<l}, u).
  \label{eq:coa_decomp}
\end{equation}
We emphasize that CoA is a \emph{practical approximation} to the
theoretical upper bound of \cref{eq:chain_rule_features}: rather than
decoding over the full item feature vector, it selectively recovers the
coarse-grained attributes most discarded by SID compression, subject to
a fixed latency overhead of $m$ additional autoregressive steps. The degree
to which this approximation narrows the feature coverage gap is quantified
by the mutual information $I(\vect{a};\, s_l \mid s_{<l}, u)$ between the
generated attributes and SID tokens---a larger mutual information
implies a greater reduction in decoding uncertainty.

This design yields two complementary benefits. First, CoA provides a
measurable reduction in generation uncertainty: the entropy difference at
layer $l$,
\begin{equation}
  \Delta H_l
  = H(s_l \mid s_{<l}, u) - H(s_l \mid \vect{a}, s_{<l}, u)
  = I(\vect{a};\, s_l \mid s_{<l}, u) \ge 0,
  \label{eq:entropy_reduction}
\end{equation}
is strictly positive whenever $\vect{a}$ and $s_l$ are not conditionally
independent---a condition that holds in practice because items sharing the
same category occupy adjacent regions in the SID semantic space, so
their token sequences are structurally correlated with coarse-grained
attributes by construction. The cumulative reduction $\sum_{l} \Delta H_l$
stabilizes the beam search trajectory; furthermore, since conditioning on
$\vect{a}$ reduces each layer's token error rate from $\epsilon_l$ to
$\epsilon_l' < \epsilon_l$, the cascading failure probability satisfies
$P'(\text{error}) = 1 - \prod_{l}(1-\epsilon_l') < 1 -
\prod_{l}(1-\epsilon_l) = P(\text{error})$, yielding end-to-end error
attenuation whenever attributes are strongly correlated with items. Second,
CoA is complementary to the cross-attention module
(\cref{sec:cross-attention}): whereas cross-attention captures user--item
interactions at the \emph{representation} level by encoding user preference
context into the decoder's hidden states, CoA operates at the
\emph{decoding} level by providing explicit item-side semantic context at
each generation step. Together, they inject \emph{who the user is} and
\emph{what kind of item} is being generated into the same generative
trajectory, enabling \textsc{UniRec} to jointly leverage user preference
structure and item attribute structure within a single framework.

In summary, CoA provides the first information-theoretic
characterization of decoding uncertainty in GR. Grounded in the
theoretical upper bound relating discriminative and generative recommendation, 
it operationalizes a principled and latency-aware approximation: 
selectively recovering the item-side feature signals most discarded by 
SID compression, with the degree of gap reduction governed by
$I(\vect{a}; s_l \mid s_{<l}, u)$ and validated empirically through
measurable per-step entropy reduction and end-to-end error attenuation.

\subsection{Conditional Decoding Context}
\label{sec:cdc}

CDC augments the decoding process with two complementary mechanisms: Task-Conditioned BOS tells the model \emph{what task} it is solving, while Content Summary provides compact \emph{combinatorial interaction features} of previously decoded tokens.

\paragraph{Task-Conditioned BOS}
A unified GR model typically needs to handle multiple objectives simultaneously---different recommendation scenarios may require optimizing for distinct user behaviors or business goals. For instance, click prediction prioritizes immediate engagement while purchase prediction focuses on conversion intent; a main feed scenario emphasizes broad interest exploration while a search scenario requires query-conditioned relevance; similarly, domestic markets may emphasize local brands whereas international markets require cross-border product discovery. Training separate models for each scenario is prohibitively expensive, while naively sharing parameters without explicit conditioning leads to objective interference, manifesting as unstable predictions and degraded performance. Task-Conditioned BOS addresses this by replacing the fixed BOS token with a learnable embedding conditioned on task context $c_{\text{task}}$, which jointly encodes the behavioral objective and recommendation scene:
\begin{equation*}
\begin{aligned}
  c_{\text{task}} \in\;
  &\underbrace{\{\text{click},\,\text{purchase},\,\text{cart},\,\text{cross-border},\,\ldots\}}_{\text{behavioral objective}} \\
  \times\;
  &\underbrace{\{\text{main feed},\,\text{search},\,\text{similar items},\,\text{flash sale},\,\ldots\}}_{\text{recommendation scene}}
\end{aligned}
\end{equation*}
This allows the model to learn distinct generation distributions for different conditions without altering the decoding architecture. By injecting task-specific signals at the initial state, Task-Conditioned BOS steers the entire generation trajectory: it biases token distributions toward contextually relevant semantic regions, maintains consistent intent throughout the decoding chain, and enables the model to dynamically adapt its generation strategy based on the recommendation scenario.

\paragraph{Content Summary}
In hierarchical decoding, the model has access to the autoregressive prefix, but individual token embeddings fail to capture their \emph{joint combinatorial semantics}. The relationship between tokens in the decoding path is not one-to-one: the same SID token may carry different semantic meanings when paired with different attributes or parent tokens, and meaningful item patterns often emerge from specific combinations rather than individual tokens. Recent work on SIDs~\cite{better_generalization} has demonstrated that subword-based decomposition of token sequences---grouping codes into n-grams or learnable subpieces---significantly improves ranking model generalization. Meanwhile, existing GR methods rely on self-attention mechanisms to implicitly learn token interactions during training, which suffers from limited expressiveness for capturing explicit combinatorial patterns.

To address this, we explicitly model Cartesian product interactions along the decoding path. However, explicitly storing all token pair combinations leads to parameter explosion (e.g., $4000^2 \approx 16$M entries for two 4000-vocabulary layers). We instead employ a Bloom filter-inspired hash projection with multiple hash functions sharing a single embedding table. For the $t$-th decoding step, we compute the content summary $\vect{c}_t$ as:
\begin{equation}
    \vect{c}_t = \bigoplus_{i=1}^{M} \vect{E}_{\text{hash}}\Bigl(H_i(\text{path}_{<t}) \bmod S\Bigr),
    \label{eq:content_summary}
\end{equation}
where $\bigoplus$ denotes concatenation, $\text{path}_{<t}$ denotes all tokens decoded before step $t$, $\{H_i\}_{i=1}^{M}$ are $M$ hash functions with different combination strategies, $\vect{E}_{\text{hash}} \in \mathbb{R}^{S \times d_{\text{hash}}}$ is a shared embedding table, and $S$ is the hash table size determined by $S = \lfloor (\prod_{i=1}^{n} V_i)^{2/(n+1)} \rfloor$ for an $n$-way Cartesian product with vocabularies $\{V_i\}$. By using multiple hash functions with different combination strategies on a shared embedding table, items with similar attribute-SID patterns naturally produce overlapping hash signatures, enabling the model to generalize across related item combinations. This formulation efficiently injects layer-specific combinatorial context with manageable parameters ($M \times S \times d_{\text{hash}}$) while preserving the semantic structure essential for generalization.

\subsection{Generation Model and Training}
\label{sec:model}

\textsc{UniRec} uses a Decoder-Only backbone with Cross-Attention to the user behavior sequence and per-step Rank Heads, as illustrated in Figure~\ref{fig:cag_overview}.

\subsubsection{Input Feature Modeling}

The model input comprises three parts: user static profile, behavior sequence, and SID-level multimodal features.

\paragraph{Static Profile Features}
Sparse fields (user ID, demographics, context features) are embedded and processed:
\begin{equation}
\mathbf{h}_{\text{static}} = \mathrm{RMSNorm}([\vect{e}_{\mathrm{uid}} \oplus \vect{e}_{\mathrm{ctx}} \oplus \cdots]) \in \mathbb{R}^{d_{\text{static}}}.
\end{equation}

\paragraph{Behavior Sequence Features}
User click behaviors are organized chronologically. Each behavior's item-side attributes (item, shop, category, etc.) are processed as:
\begin{equation}
\vect{h}_i = \mathrm{Linear}(\mathrm{RMSNorm}([\vect{e}_{\mathrm{item}_i} \oplus \vect{e}_{\mathrm{shop}_i} \oplus \vect{e}_{\mathrm{cate}_i} \oplus \cdots])) \in \mathbb{R}^{d_{\text{model}}},
\end{equation}
forming the behavior sequence $\mathbf{H}_{\text{seq}} = \{\vect{h}_1, \ldots, \vect{h}_T\} \in \mathbb{R}^{T \times d_{\text{model}}}$.

\paragraph{SID-Level Multimodal Features}
Multi-layer SIDs $\{s_{0}, s_{1}, s_{2}\}$ derived from multimodal content alignment are embedded and fused with their corresponding item embeddings before projection to ensure temporal alignment with the behavior sequence:
\begin{equation}
    \mathbf{H}_{\text{mm}} = \{\vect{h}_1^{\mathrm{mm}}, \ldots, \vect{h}_{L_{\text{mm}}}^{\mathrm{mm}}\} \in \mathbb{R}^{L_{\text{mm}} \times d_{\text{model}}},
\end{equation}
where each $\vect{h}_j^{\mathrm{mm}} = \mathrm{RMSNorm}(\mathrm{Linear}([\vect{e}^{(0)}_j \oplus \vect{e}^{(1)}_j \oplus \vect{e}^{(2)}_j]))$.
These are appended to the sequence: $\mathbf{H}_{\text{seq}} \gets [\mathbf{H}_{\text{seq}} ; \mathbf{H}_{\text{mm}}] \in \mathbb{R}^{(T+L_{\text{mm}}) \times d_{\text{model}}}$.

\paragraph{Aggregated Representation}
The static profile and pooled behavior features are combined:
\begin{equation}
\mathbf{h}_{\text{agg}} = \mathrm{RMSNorm}( [\mathbf{h}_{\text{static}} \oplus \mathrm{Pool}(\mathbf{H}_{\text{seq}})]) \in \mathbb{R}^{d_{\text{agg}}}.
\end{equation}

\subsubsection{Cross-Attention Conditioning}
\label{sec:cross-attention}

Cross-Attention decouples the user behavior context from the token decoding process. In this module, the behavior sequence $\mathbf{H}_{\text{seq}}$ serves as the static Key-Value pairs, while the decoding-side sequence---comprising the task prompt, attributes, and SID tokens---acts as the Query.

Specifically, let $\mathbf{e}_{\text{BOS}}$ be the Task-Conditioned BOS embedding (dependent on $c_{\text{task}}$), $\mathbf{e}_{\text{attr}}$ the embeddings of the CoA tokens $[a_1, \ldots, a_m]$, and $\{\mathbf{e}_0, \mathbf{e}_1, \mathbf{e}_2\}$ the embeddings of the hierarchical SID tokens. The query sequence $\mathbf{Q}$ is constructed by concatenating these embeddings with positional encodings:
\begin{equation}
\mathbf{Q}^{(0)} = \text{PosEncoding}([\mathbf{e}_{\text{BOS}} \oplus  \mathbf{e}_{\text{attr}} \oplus \mathbf{e}_0 \oplus  \mathbf{e}_1\oplus  \mathbf{e}_2]) \in \mathbb{R}^{(1+m+L) \times d_{\text{model}}},
\end{equation}
where the total query length \(1+m+L\) corresponds to the full decoding path. The Key and Value matrices are derived from the unified behavior sequence $\mathbf{H}_{\text{seq}}$, also augmented with positional information:
\begin{equation}
\mathbf{K} = \mathbf{V} = \text{PosEncoding}(\mathbf{H}_{\text{seq}}) \in \mathbb{R}^{(T + L_{\text{mm}}) \times d_{\text{model}}}.
\end{equation}

The query sequence is processed through $D$ successive cross-attention layers. To maintain stable gradients and effective information flow, each layer $d$ employs Pre-Norm and residual connections. The operations for layer $d$ are formulated as:
\begin{align}
\mathbf{Q}^{(d)}_{\text{mid}} &= \mathbf{Q}^{(d-1)} + \text{GatedCrossAttn}(\text{RMSNorm}(\mathbf{Q}^{(d-1)}), \mathbf{K}, \mathbf{V}), \\
\mathbf{Q}^{(d)} &= \mathbf{Q}^{(d)}_{\text{mid}} + \text{MMoE-FFN}(\text{RMSNorm}(\mathbf{Q}^{(d)}_{\text{mid}})),
\end{align}
where $\text{GatedCrossAttn}$ utilizes a learnable gating parameter $\gamma$ to modulate the contribution of the behavior context:
\begin{equation}
\text{GatedCrossAttn}(\mathbf{Q}, \mathbf{K}, \mathbf{V}) = \gamma \cdot \text{Softmax}\left(\frac{\mathbf{Q}\mathbf{K}^\top}{\sqrt{d_{\text{model}}}}\right)\mathbf{V}.
\end{equation}
Each expert within the $\text{MMoE-FFN}$ utilizes SwiGLU activations to capture complex task-specific patterns. After $D$ layers of refinement, the final output sequence is normalized:
\begin{equation}
\mathbf{Q}_{\text{out}} = \text{RMSNorm}(\mathbf{Q}^{(D)}) \in \mathbb{R}^{(1+m+L) \times d_{\text{model}}}.
\end{equation}

\subsubsection{Hierarchical Rank Head}

\textsc{UniRec} employs a dedicated Rank Head for each decoding step $t \in \{1, \ldots, m+L\}$. The input $\mathbf{x}_t$ combines: (1) cross-attention output $\mathbf{q}_t = \mathbf{Q}_{\text{out}}[t, :]$, (2) prefix embeddings $\mathbf{e}_{\text{prefix}}$ of tokens decoded before step $t$, (3) Content Summary $\mathbf{c}_t$ from \cref{eq:content_summary}, and (4) aggregated representation $\mathbf{h}_{\text{agg}}$.

The Rank Head processes $\mathbf{x}_t = [\mathbf{q}_t \oplus \mathbf{e}_{\text{prefix}} \oplus \mathbf{c}_t \oplus \mathbf{h}_{\text{agg}}]$ through SENet~\cite{hu2018senet} and MaskNet~\cite{wang2021masknet} to produce:
\begin{equation}
  p(s_t \mid s_{<t}, u, c_{\text{task}}) = \mathrm{Softmax}(g^{(t)}(\mathbf{x}_t)),
  \label{eq:layer_prob}
\end{equation}
where $g^{(t)}(\cdot)$ denotes the Rank Head at step $t$. The output vocabulary is task-specific: attribute domains for $t \le m$, and SID layer $\mathcal{V}_{t-m-1}$ for $t > m$.

\subsubsection{Training Objectives}

The generation module is trained via NTP with teacher forcing. The target item is represented as $\mathbf{s}^\star = (a^\star_1, \ldots, a^\star_m, s^\star_0, \ldots, s^\star_{L-1})$ comprising $m$ attribute tokens and $L$ SID tokens. The training loss is:
\begin{equation}
  \mathcal{L}_{\mathrm{NTP}} = -\sum_{t=1}^{m+L} \alpha_i \cdot \log p_\theta(s^\star_t \mid \mathbf{s}^\star_{<t}, u, c_{\text{task}}),
  \label{eq:ntp_hierarchical}
\end{equation}
where $\alpha_i$ weights samples by engagement (e.g., click, conversion), and $p_\theta$ is produced by Rank Head $g^{(t)}$ from \cref{eq:layer_prob}. AdamW is employed as the optimizer to minimize the objective function.

\subsection{Business Objective and User Preference Alignment}
\label{sec:alignment}

NTP trains \textsc{UniRec} to model the exposure 
distribution, but optimizes for distribution matching rather than 
business objectives. To bridge this gap, we introduce a unified 
alignment framework: RFT 
reformulates the NTP objective by reweighting training samples 
according to continuous business value estimates (e.g., GMV, watch 
time); DPO further 
injects discrete behavioral preference signals (purchase, click, 
exposure) via contrastive pair optimization. The two are jointly 
optimized in a single training step:
\begin{equation}
    \mathcal{L} = \mathcal{L}_{\text{RFT}} + 
    \lambda_{\text{DPO}}\,\mathcal{L}_{\text{DPO}},
    \label{eq:total_loss}
\end{equation}
where $\mathcal{L}_{\text{RFT}}$ is the business-value-reweighted 
NTP objective (\cref{eq:rft_loss}) and $\lambda_{\text{DPO}}$ 
controls the relative contribution of the preference signal.

\subsubsection{Reward-Driven Fine-tuning}
\label{sec:rft}

RFT reformulates NTP by upweighting samples whose interacted items 
carry higher estimated business value. For each training sample 
$(u_i, x_i)$, we define a composite reward:
\begin{equation}
    R(u_i, x_i) = \mathcal{F}(\{\hat{y}_k(u_i, x_i)\}_{k=1}^{N_{\text{obj}}}),
    \label{eq:reward}
\end{equation}
where $\{\hat{y}_k\}$ denote predicted engagement metrics (e.g., 
watch time, conversion probability) and $\mathcal{F}$ aggregates 
them according to business priorities. We normalize sample 
advantages within each batch $\mathcal{B}$ to stabilize gradients 
across varying reward scales:
\begin{equation}
    A_i = R(u_i, x_i) - \frac{1}{|\mathcal{B}|} 
    \sum_{j \in \mathcal{B}} R(u_j, x_j),
\end{equation}
\begin{equation}
    \hat{A}_i = \frac{A_i}{\sigma_A + \epsilon}, \quad 
    \tilde{A}_i = \text{clip}(\hat{A}_i, -c_{\text{clip}}, 
    c_{\text{clip}}),
\end{equation}
where $\sigma_A = \sqrt{\frac{1}{|\mathcal{B}|} \sum_{j \in 
\mathcal{B}} A_j^2}$ is the batch root mean square, $\epsilon$ 
prevents division by zero, and $c_{\text{clip}}$ suppresses 
outliers. The reweighted training objective is:
\begin{equation}
    \mathcal{L}_{\text{RFT}} = -\sum_{i \in \mathcal{B}} 
    \sum_{t=1}^{m+L} (1 + \lambda \tilde{A}_i) \cdot \alpha_i 
    \cdot \log p_\theta(s^\star_{i,t} \mid \mathbf{s}^\star_{i,<t}, 
    u_i, c_{\text{task}}),
    \label{eq:rft_loss}
\end{equation}
where $\lambda > 0$ controls reweighting strength, and $\alpha_i$ 
is proportional to the GMV value of the interacted item, such that 
high-value transactions contribute more strongly to the training 
signal. When $\tilde{A}_i > 0$ the model amplifies learning from 
high-reward samples; when $\tilde{A}_i < 0$ it suppresses 
low-performing patterns.

\subsubsection{Preference Alignment via Direct Preference 
Optimization}
\label{sec:dpo}

DPO complements RFT by directly contrasting item pairs according 
to observed behavioral outcomes. For each item $x$ exposed under 
request context $u$, we define a behavioral preference level:
\begin{equation}
    \mathcal{R}(x) = \begin{cases} 
    2 & \text{if } x \text{ is purchased} \\ 
    1 & \text{if } x \text{ is clicked} \\ 
    0 & \text{if } x \text{ is exposed only} 
    \end{cases}
\end{equation}
Items are ordered as $x_i \succ x_j$ if $\mathcal{R}(x_i) > 
\mathcal{R}(x_j)$, or by exposure rank when behaviors are equal. 
Preference pairs are organized by request: within each batch, 
all items exposed under the same request context $u$ are grouped, 
and pairs $(x_i, x_j)$ satisfying $x_i \succ x_j$ are sampled 
from each group to construct:
\begin{equation}
    \mathcal{D} = \{(u, x_i, x_j) \mid x_i \succ x_j,\ 
    x_i, x_j \in \mathcal{E}_u\},
\end{equation}
where $\mathcal{E}_u$ denotes the set of items exposed under 
request $u$. The DPO objective is:
\begin{equation}
    \mathcal{L}_{\text{DPO}} = -\mathbb{E}_{(u,\, y_w,\, y_l) 
    \sim \mathcal{D}} \left[ \log \sigma \!\left( \beta \left( 
    \log \frac{\pi_\theta(y_w \mid u)}{\pi_{\text{ref}}(y_w \mid u)}
    - \log \frac{\pi_\theta(y_l \mid u)}{\pi_{\text{ref}}(y_l \mid 
    u)} \right) \right) \right],
\end{equation}
where $\pi_{\text{ref}}$ is the reference model and $\beta=0.1$ 
controls preference margin sharpness.

\paragraph{Layer-wise Stop Gradient.}
To preserve the stability of prefix predictions that serve as 
conditioning context for subsequent generation, we apply 
stop-gradient to all decoding steps except the final SID layer, 
whose gradients are allowed to flow. Formally:
\begin{equation}
    \log \pi_\theta(y \mid u) =
    \underbrace{\sum_{t=1}^{L-1} \operatorname{sg}\!\left[\log 
    p_\theta\!\left(s_t^\star \mid s_{<t}^\star, u\right)\right]}
    _{\text{frozen prefix}}
    +
    \underbrace{\log p_\theta\!\left(s_{L}^\star \mid s_{<L}^\star, 
    u\right)}_{\text{DPO target layer}},
\end{equation}
where $\operatorname{sg}[\cdot]$ denotes the stop-gradient 
operator. This ensures DPO exclusively updates the final-layer 
Rank Head while leaving all prefix predictions intact.

\section{Experiments}

We conduct extensive offline and online experiments to validate the effectiveness of \textsc{UniRec}. Our offline evaluation demonstrates improvements in token prediction accuracy and retrieval quality across multiple metrics, while online A/B testing confirms significant gains in business objectives including GMV and user engagement.


\subsection{Offline Evaluation}

\subsubsection{Experimental Setup}

\paragraph{Dataset}
We conduct offline experiments on Shopee’s e-commerce platform. The dataset is constructed from real production logs over 9 consecutive days in the platform's main feed recommendation scenario, where users browse a continuous vertical scroll of products. The dataset contains billions of user interaction records, covering diverse behaviors including impressions, clicks, add-to-cart actions, and purchase conversions.

\paragraph{Evaluation Metrics}
We evaluate model performance using the following metrics:
\begin{itemize}[leftmargin=*,noitemsep,topsep=3pt]
    \item \textbf{Token Hit Ratio@3}: During teacher-forcing training, we measure the fraction of ground-truth tokens that appear in the model's top-3 predictions at each decoding step. This metric reflects how well the model learns the token distribution.

    \item \textbf{BS Hit Ratio@K}: At inference time, we perform beam search and measure whether the target item appears in the top-$K$ generated candidates. We report Hit Ratio at $K\in\{50,100,200\}$. Given that purchase conversions are the most valuable interactions for e-commerce, we additionally report BS Hit Ratio@$K$ specifically on order samples (the subset of samples corresponding to successful purchases).
\end{itemize}

\paragraph{Model Configuration}
For Capacity-constrained SIDs, we use a 3-layer hierarchy 
with codebook size $K=4000$ per layer and tolerance $\tau=1.05$. 
For CoA, we decode the item's category hierarchy 
(L2 $\to$ L3)\footnote{Categories are organized hierarchically as 
L1 (top-level) $\supset$ L2 (mid-level) $\supset$ L3 
(fine-grained), where each level provides progressively finer item 
categorization.} before SID tokens. Content Summary employs 
hash-based feature crossing with $M=3$ hash functions over feature 
index pairs $(x,y)$: $H_1(x,y) = x + y$, $H_2(x,y) = x \cdot y$, 
and $H_3(x,y) = p_1 x + p_2 y$ where $p_1, p_2$ are prime numbers. 
The hash embedding dimension is $d_{\text{hash}} = 64$, and 
Cartesian products include: (L2, $s_0$), (L2, $s_1$), (L3, $s_0$), 
(L3, $s_1$), and ($s_0$, $s_1$). The model dimension is 
$d_{\text{model}}=256$, with behavior sequence length $T=200$ and 
multimodal SID sequence length $L_{\text{mm}}=100$. The 
cross-attention module uses 3 layers with 8 attention heads per 
layer. The MMoE-FFN employs 4 experts with SwiGLU activation and 
hidden dimension $4d_{\text{model}}$. For Business Objective 
Alignment, we use GMV as the business reward signal for RFT, with 
loss weights $\lambda_{\text{RFT}}:\lambda_{\text{DPO}} = 20:3$; 
DPO is configured with $\beta=0.1$ and preference pairs are 
constructed by grouping exposed items per request and sampling 
within each group. The model is trained with the AdamW optimizer using a 
learning rate of $3 \times 10^{-4}$.

\subsubsection{Baseline Comparison}

We compare \textsc{UniRec} against the following methods:
\begin{itemize}[leftmargin=*,noitemsep,topsep=3pt]
    \item \textbf{SASRec}~\cite{kang2018sasrec}: A self-attention based sequential recommendation model that serves as a strong discriminative retrieval baseline. We adapt SASRec to our SID-based evaluation by using its learned representations for nearest-neighbor retrieval over the SID codebook.
    \item \textbf{TIGER}~\cite{tiger}: An Encoder-Decoder generative retrieval architecture where the encoder processes user behavior sequences and the decoder autoregressively generates item identifiers.
    \item \textbf{OneRec-V2}~\cite{onerec,zhou2025onerecv2}: A Decoder-Only generative retrieval architecture with cross-attention to user behavior sequences for autoregressive item generation.
\end{itemize}

Table~\ref{tab:baseline} presents the performance comparison on all samples. \textsc{UniRec} achieves substantial improvements across all metrics, outperforming the strongest generative baseline (OneRec-V2) by +9.9 percentage points at HR@50 (+22.6\% relative gain), +9.5 points at HR@100 (+18.2\% relative gain), and +10.1 points at HR@200 (+17.2\% relative gain). These gains demonstrate the effectiveness of our unified framework design.

\paragraph{Performance on Order Samples}
Given the critical importance of purchase conversions in e-commerce, we separately evaluate Hit Ratio on order samples (Table~\ref{tab:baseline}, right half). \textsc{UniRec} demonstrates even stronger performance on this high-value subset, outperforming OneRec-V2 by +9.0 points at HR@50, +10.3 points at HR@100, and +11.5 points at HR@200. This amplified performance on purchase conversions stems from \textsc{UniRec}'s ability to jointly model multiple user actions inherent to e-commerce scenarios. Our Task-Conditioned BOS mechanism enables the model to distinguish between diverse behavioral objectives---click, add-to-cart, purchase, and cross-border transactions---while RFT reweights samples based on actual business value (GMV).

\begin{table*}[!ht]
\centering
\caption{Hit Ratio on all samples and order (purchase) samples.}
\label{tab:baseline}
\begin{tabular}{lcccccc}
\toprule
 & \multicolumn{3}{c}{\textbf{All Samples}} & \multicolumn{3}{c}{\textbf{Order Samples}} \\
\cmidrule(lr){2-4}\cmidrule(lr){5-7}
\textbf{Method}
  & \textbf{HR@50} & \textbf{HR@100} & \textbf{HR@200}
  & \textbf{HR@50} & \textbf{HR@100} & \textbf{HR@200} \\
\midrule
SASRec            & 0.421 & 0.489 & 0.556 & 0.548 & 0.631 & 0.709 \\
TIGER (0.05B)     & 0.437 & 0.508 & 0.578 & 0.567 & 0.652 & 0.731 \\
OneRec-V2 (0.05B) & 0.438 & 0.523 & 0.587 & 0.582 & 0.671 & 0.752 \\
\midrule
\textsc{UniRec} (0.05B)
  & \textbf{0.537} & \textbf{0.618} & \textbf{0.688}
  & \textbf{0.672} & \textbf{0.774} & \textbf{0.867} \\
\bottomrule
\end{tabular}
\end{table*}

\begin{table*}[ht!]
\centering
\caption{Ablation of CoA configurations.}
\label{tab:ablation_coa}
\begin{tabular}{lccccccccc}
\toprule
\textbf{Attribute Chain}
  & \textbf{L1@3} & \textbf{L2@3} & \textbf{L3@3}
  & \textbf{$s_0$@3} & \textbf{$s_1$@3} & \textbf{$s_2$@3}
  & \textbf{HR@50} & \textbf{HR@100} & \textbf{HR@200} \\
\midrule
Direct SID (no attributes)
  & --- & --- & --- & 0.314 & 0.752 & 0.952 & 0.458 & 0.532 & 0.595 \\
L1 $\to$ SID
  & 0.899 & --- & --- & 0.488 & 0.776 & 0.961 & 0.471 & 0.569 & 0.622 \\
L2 $\to$ SID
  & --- & 0.755 & --- & 0.591 & 0.789 & 0.963 & 0.515 & 0.597 & 0.664 \\
L3 $\to$ SID
  & --- & --- & 0.694 & 0.654 & 0.801 & 0.967 & 0.517 & 0.601 & 0.670 \\
L2 $\to$ L3 $\to$ SID
  & --- & 0.757 & 0.962 & 0.682 & 0.821 & 0.971
  & \textbf{0.537} & \textbf{0.618} & \textbf{0.688} \\
\bottomrule
\end{tabular}
\end{table*}

\begin{table}[ht]
\centering
\caption{%
  Ablation and scaling results.
  \textsc{UniRec} (Full Model) is the shared reference for all sub-tables.
  \textbf{(a)} SID construction ablation.
  \textbf{(b)} CDC component ablation.
  \textbf{(c)} Model scaling ($d_{\text{model}}$);
  $^\dagger$denotes the default setting, identical to Full Model.
}
\label{tab:ablation}
\begin{tabular}{lccc}
\toprule
\textbf{Configuration} & \textbf{HR@50} & \textbf{HR@100} & \textbf{HR@200} \\
\midrule
\textsc{UniRec} (Full Model)
  & \textbf{0.537} & \textbf{0.618} & \textbf{0.688} \\
\midrule
\multicolumn{4}{l}{\textit{(a) SID Construction}} \\[2pt]
\quad RQ-KMeans         & 0.481 & 0.558 & 0.627 \\
\midrule
\multicolumn{4}{l}{\textit{(b) Conditional Decoding Context}} \\[2pt]
\quad w/o Task-Cond. BOS   & 0.493 & 0.570 & 0.638 \\
\quad w/o Content Summary & 0.485 & 0.565 & 0.637 \\
\midrule
\multicolumn{4}{l}{\textit{(c) Model Scaling ($d_{\text{model}}$)}} \\[2pt]
\quad 64                & 0.397 & 0.459 & 0.514 \\
\quad 128               & 0.477 & 0.551 & 0.615 \\
\quad 256$^\dagger$     & 0.537 & 0.618 & 0.688 \\
\quad 512               & 0.557 & 0.639 & 0.712 \\
\bottomrule
\end{tabular}
\end{table}

\subsubsection{Ablation Studies}

We ablate three core components of \textsc{UniRec}: Capacity-Constrained SID construction, and the two CDC components (Task-Conditioned BOS and Content Summary). All variants are compared against the Full Model reference in Table~\ref{tab:ablation}.

\paragraph{Capacity-Constrained SID}

\begin{figure}[ht]
    \centering
    \includegraphics[width=\linewidth]{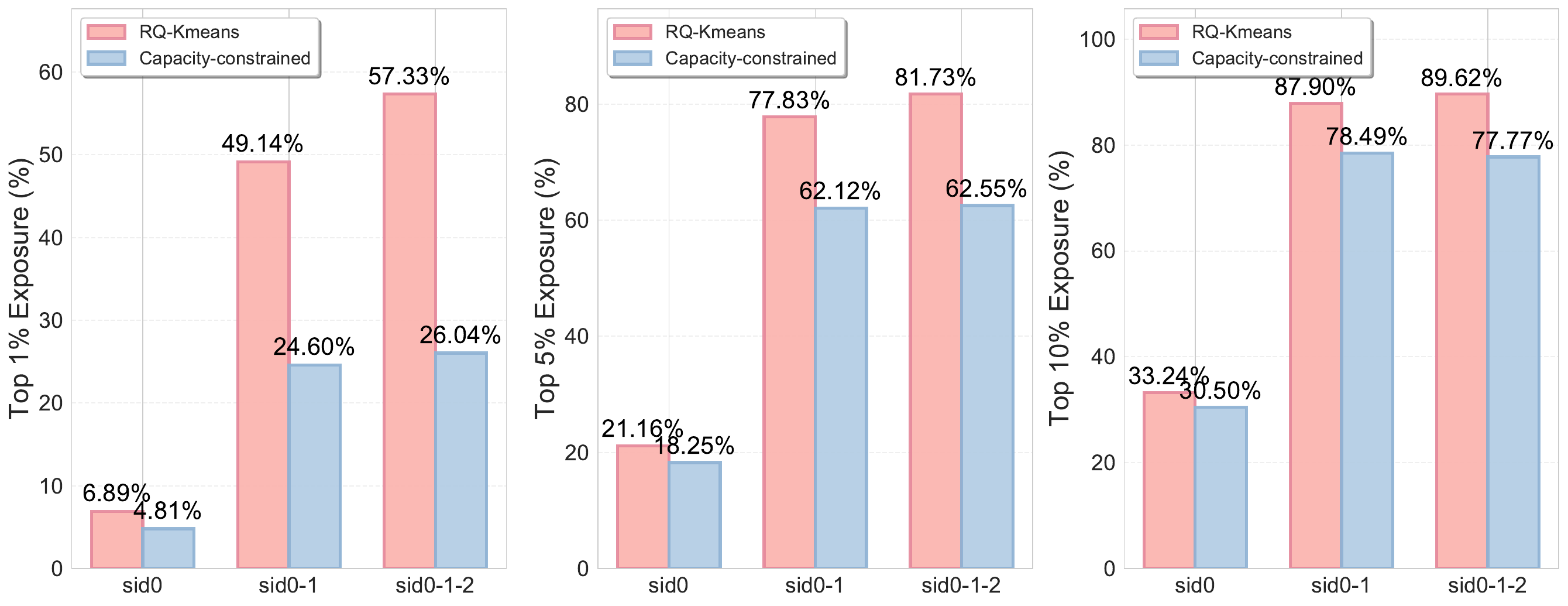}
    \caption{%
      Matthew-effect analysis of SID token exposure concentration.
      Lower values indicate more balanced codebook utilization.
      At the full three-level hierarchy (sid0-1-2), the top-1\% tokens under
      RQ-KMeans capture 57.33\% of all exposures, whereas Capacity-Constrained
      SIDs reduce this to 26.04\%.}
    \label{fig:matthew_effect}
\end{figure}

Replacing Capacity-Constrained SIDs with standard RQ-KMeans leads to consistent degradation across all metrics (Table~\ref{tab:ablation}(a)), with HR@50 dropping from 0.537 to 0.481 and HR@100 from 0.618 to 0.558. Beyond retrieval accuracy, we examine the \emph{codebook utilization} of the two approaches to understand the underlying cause. Figure~\ref{fig:matthew_effect} plots the fraction of total item exposures captured by the top-1\%, top-5\%, and top-10\% most-used tokens at each level of the SID hierarchy (sid0, sid0-1, and sid0-1-2). The imbalanced distribution in RQ-KMeans forces the decoder to disproportionately allocate its probability mass to a small set of over-represented codes, degrading generalization to less common items. Capacity-Constrained SIDs enforce a near-uniform load across the codebook through the occupancy tolerance $\tau$, producing a more faithful discrete representation for the long-tail item distribution prevalent in e-commerce.

\paragraph{Task-Conditioned BOS}

Removing Task-Conditioned BOS leads to a relative drop of 7.8\% at HR@100 (Table~\ref{tab:ablation}(b)), as the model loses the ability to distinguish between task-specific objectives.

To further validate the benefit of Task-Conditioned BOS in a \emph{multi-scene} setting, we train on the joint sample pool of three production scenarios and compare three conditions (Table~\ref{tab:multiscene}):
\begin{enumerate}[leftmargin=*,noitemsep,topsep=2pt]
    \item Single-scene: independent model trained and evaluated per scene.
    \item Multi-scene: one model trained on all scenes jointly, without scene-differentiating prompts.
    \item Multi-scene + Task-Cond.\ BOS: joint training with scene-specific task-conditioned BOS embeddings.
\end{enumerate}

\begin{table}[ht]
\centering
\caption{Multi-scene effectiveness of Task-Conditioned BOS.}
\label{tab:multiscene}
\begin{tabular}{lccc}
\toprule
\textbf{Configuration} & \textbf{HR@50} & \textbf{HR@100} & \textbf{HR@200} \\
\midrule
Single-scene                & 0.386 & 0.470 & 0.540 \\
Multi-scene                 & 0.395 & 0.486 & 0.570 \\
Multi-scene + Task-Cond.\ BOS   & \textbf{0.400} & \textbf{0.510} & \textbf{0.590} \\
\bottomrule
\end{tabular}
\end{table}

Joint training with multi-scene data already improves over per-scene models (+0.9/+1.6/+3.0 points at HR@50/100/200), demonstrating the benefit of cross-scene knowledge transfer. Introducing Task-Conditioned BOS provides a further consistent gain at all cutoffs, confirming that scene-specific conditioning is necessary to resolve conflicts between heterogeneous behavioral distributions when training on mixed data.

\paragraph{Content Summary}

Removing Content Summary causes a larger relative degradation of 8.6\% at HR@100 (Table~\ref{tab:ablation}(b)), indicating that explicitly modeling Cartesian-product feature interactions along the decoding path is critical for capturing combinatorial signals that sequence modeling alone cannot efficiently represent. Each hash function $H_i$ maps a feature index pair $(x,y)$ to a shared embedding table, enabling the model to encode cross-feature correlations (e.g., category $\times$ SID token) as a lightweight inductive bias injected at the prompt layer. The full model combining both Task-Conditioned BOS and Content Summary achieves the best performance, confirming that task conditioning and combinatorial feature crossing address orthogonal challenges.

\subsubsection{Model Scaling Analysis}

We investigate how model capacity affects performance by scaling $d_{\text{model}}$ while fixing all other hyperparameters. Table~\ref{tab:ablation}(c) shows that HR@50, HR@100, and HR@200 improve consistently as $d_{\text{model}}$ increases from 64 to 512. The scaling trend suggests that \textsc{UniRec} benefits from additional capacity, leaving room for further improvements with larger architectures.

\subsection{Online A/B Testing}
We deploy \textsc{UniRec} in production on Shopee's e-commerce platform, serving tens of millions of users daily.

\paragraph{Experimental Setup}
We conduct A/B experiments covering both main feed and landing 
page scenarios. The control groups use the baseline discriminative 
multi-stage recommender system, while treatment groups deploy 
\textsc{UniRec}. Experiments are conducted with 20\% user traffic 
allocation per bucket to ensure statistical significance.

\paragraph{Experiment Results}
Table~\ref{tab:online_main_feed} presents the results across both 
scenarios. Overall, \textsc{UniRec} achieves substantial 
improvements across all key metrics: +5.37\% PVCTR, +4.76\% 
orders, +5.60\% GMV. The main feed segment shows consistent gains with 
+5.37\% PVCTR, +4.27\% orders, and +5.42\% GMV. 
The landing page segment delivers +5.78\% orders and +6.19\% GMV.

\begin{table}[ht]
\centering
\caption{Online A/B test results.}
\label{tab:online_main_feed}
\small
\begin{tabular}{@{}lccc@{}}
\toprule
\textbf{Metric} & \textbf{Overall} & \textbf{Feed} & \textbf{Landing} \\
\midrule
Total Orders  & +4.76\% & +4.27\% & +5.78\% \\
GMV           & +5.60\% & +5.42\% & +6.19\% \\
Page-view CTR & +5.37\% & +5.37\% & ---     \\
\bottomrule
\end{tabular}
\end{table}

\paragraph{Key Observations}
Across all experiments, several consistent patterns emerge: 
(1) \textsc{UniRec} delivers stronger relative gains in PVCTR and engagement metrics compared to conversion metrics, 
indicating improved content relevance and user satisfaction. 
(2) The system maintains competitive latency (110ms end-to-end) 
comparable to a single ranking model, while the traditional 
pipeline requires 266ms, demonstrating the efficiency gains from 
end-to-end optimization.

\section{Conclusion}
In this paper, we present \textsc{UniRec}, a GR 
framework that addresses the fundamental structural 
gap between generative and discriminative recommendation. We first 
establish a theoretical foundation via Bayes' theorem, proving 
that the long-assumed performance gap between the two paradigms 
arises from feature coverage rather than a fundamental modeling 
asymmetry. This motivates CoA, a 
speculate-then-refine paradigm that recovers item-side feature 
signals within the generative trajectory and yields provable 
per-step entropy reduction and end-to-end error attenuation. 
Beyond CoA, Capacity-constrained SID suppresses the hereditary 
Matthew effect in SID token distributions through 
exposure-weighted residual quantization, and CDC stabilizes 
multi-scenario decoding through Task-Conditioned BOS and 
combinatorial Content Summaries. A joint RFT and DPO framework 
further aligns the model with heterogeneous business objectives. 
Extensive offline experiments show that \textsc{UniRec} 
outperforms the strongest baseline by +22.6\% relative HR@50 on 
all samples and +15.5\% on high-value order samples. Deployed on 
Shopee's e-commerce platform, online A/B tests 
further confirm significant gains in PVCTR (+5.37\%), orders 
(+4.76\%), and GMV (+5.60\%), validating its effectiveness in 
complex real-world settings.

\appendix

\bibliographystyle{ACM-Reference-Format}
\bibliography{references}

\end{document}